\documentclass[twocolumn,noshowpacs,amsmath,amssymb,prl]{revtex4}

\usepackage{cases}

\usepackage{graphicx} 
\newcommand{\maximumfigurewidth}{0.44\textwidth}
\newcommand{\figurewidth}{0.42\textwidth}

\renewcommand{\epsilon}{\varepsilon}

\begin{document}
 
\title{Confinement-driven translocation of a flexible polymer}

\author{Angelo Cacciuto}
\author{Erik Luijten}
\email{luijten@uiuc.edu}
\affiliation{Department of Materials Science and Engineering and
Frederick Seitz Materials Research Laboratory, University of Illinois at
Urbana-Champaign, Urbana, Illinois 61801, U.S.A.}

\begin{abstract} 
  We consider the escape of a flexible, self-avoiding polymer chain out
  of a confined geometry. By means of simulations, we demonstrate that
  the translocation time can be described by a simple scaling law that
  exhibits a nonlinear dependence on the degree of polymerization and
  that is sensitive to the nature of the confining geometry. These
  results contradict earlier predictions but are in agreement with
  recently confirmed geometry-dependent expressions for the free energy
  of confinement.
\end{abstract}

\date{\today}

\maketitle 

Translocation through a nanopore is one of the fundamental biological
mechanisms through which long molecules can be exchanged between
different regions compartmentalized by biological
membranes~\cite{alberts02}. Examples of this phenomenon include the
injection into host cells of DNA packed inside virus
capsids~\cite{alberts02} and the transport of proteins through
biological membranes. Furthermore, pioneering experiments have
demonstrated that DNA can be translocated through a nanopore by means of
an external electric field, and that this event can be probed by
measuring the variation in ionic current through the
pore~\cite{kasianowicz96,akeson99,meller00}. This has opened the
prospect of creating efficient and economical DNA sequencing devices and
has resulted in a widespread theoretical and experimental interest in
polymer
translocation~\cite{sung96,park98b,lubensky99,muthukumar99,muthukumar01,%
chuang01,tian03,lansac04,kantor04,kong04,storm05}.

The passage of a flexible chain through a narrow opening involves a
large entropic barrier, so that most polymer translocation phenomena
require a driving force. A typical experimental
setup~\cite{kasianowicz96} consists of two chambers separated by an
interface. A DNA molecule is placed in one of the chambers and passes
through the interface via a small orifice, either a protein complex
embedded in a membrane or a solid state nanopore.  In this case, the
required driving force is provided by an external electric field.
However, one can also envisage the use of other forces, e.g., generated
by optical tweezers or by an osmotic pressure resulting from the
geometrical confinement of the polymer in one of the chambers.

In the study of polymer translocation, the duration of the sequential
passage of a chain through a membrane---measured from the entrance of
the first monomer into the pore---occupies a central place, because it
is one of the few dynamical parameters that is accessible to current
experiments~\cite{kasianowicz96,meller01,chen04,storm05}. In particular,
there have been various efforts to determine and to understand the
dependence of translocation time~$\tau$ on the degree of polymerization
and the magnitude of the driving force.  Sung and Park~\cite{sung96}
proposed to treat polymer translocation as a diffusion process across a
free energy barrier.  Muthukumar~\cite{muthukumar99} reproduced this
approach using nucleation theory, employing a corrected diffusivity.
This analysis reveals the existence of two distinct regimes.  If the
chemical potential gradient per monomer~$\Delta \mu$ is sufficiently
small, the entropic barrier dominates the diffusion process,
\begin{equation}
  \tau \sim N^2 \quad \text{for } N|\Delta\mu| \ll 1 \;,
\end{equation}
whereas stronger driving forces lead to a drift-dominated regime,
\begin{equation}
\tau \sim \frac{N}{\Delta\mu} \quad \text{for } N|\Delta\mu| > 1 \;.
\label{eq:escapetimeLinear}
\end{equation}

Chuang \emph{et al.}~\cite{chuang01} observed an inconsistency in the
reasoning leading to Eq.~(\ref{eq:escapetimeLinear}), since it assumes
that the polymer chain is equilibrated at all times during the
translocation, whereas the Rouse equilibration time scales as
$N^{2\nu+1}$ and hence for sufficiently long chains exceeds the
predicted translocation time.  Furthermore, it was
argued~\cite{kantor04} that the unhindered motion of a chain provides a
lower bound for the translocation time, such that
Eq.~(\ref{eq:escapetimeLinear}) should be replaced by
\begin{equation}
\tau \sim \frac{N^{1+\nu}}{\Delta\mu} \;.
\label{eq:escapetime}
\end{equation}

The difference in chemical potential of the monomers on either side of
the interface clearly is a crucial ingredient in all estimates of the
translocation time. Unlike other simulation
studies~\cite{tian03,lansac04}, here we concentrate on the situation
where this difference originates from polymer confinement, i.e., where
the chain is ejected out of a restricted geometry. This situation has
been analyzed in Refs.~\cite{park98b} and~\cite{muthukumar01}, where
numerical results were presented that agree with
Eq.~(\ref{eq:escapetimeLinear}), rather than with the corrected
prediction Eq.~(\ref{eq:escapetime}).  To confuse matters further, it
was recently established that the driving force exhibits a nontrivial
dependence on geometry~\cite{grosberg94,cacciuto06} that was not taken
into account properly in Refs.\ \cite{park98b} and~\cite{muthukumar01}.
In this Letter we resolve this contradictory situation by means of an
accurate numerical study of the escape of a polymer chain out of
\emph{different} confining geometries, in which we \emph{independently}
vary the degree of polymerization and the strength of the osmotic
driving force.

The free energy cost of confining a linear, flexible polymer within a
planar or cylindrical geometry (uniaxial or biaxial confinement,
respectively) is given by a well-known blob scaling
result~\cite{degennes79},
\begin{equation}
  \beta\Delta F \sim \left( \frac{R_G}{R} \right )^{1/\nu}
  \sim N \left( \frac{\sigma}{R} \right)^{1/\nu} \;,
\label{eq:plates}
\end{equation} 
where $R$ is the separation between the plates or the radius of the
cylinder and $R_G\sim\sigma N^{\nu}$ is the radius of gyration of the
polymer in its \emph{unconfined} state.  $\sigma$ is the size of a
monomer, $\nu\simeq 0.588$ is the Flory exponent and $\beta = 1/(k_{\rm
B}T)$, with $k_{\rm B}$ the Boltzmann constant and $T$ the absolute
temperature.  As has been argued on theoretical
grounds~\cite{grosberg94} and confirmed numerically~\cite{cacciuto06},
this prediction is \emph{invalid} for a spherical geometry (triaxial
confinement) and must be replaced by a scaling law that has a much
stronger dependence on the sphere radius~$R$,
\begin{equation}
  \beta\Delta F \sim \left( \frac{R_G}{R} \right)^{3/(3\nu-1)}
  \sim N\phi^{1/(3\nu-1)} \;,
\label{eq:sphere}
\end{equation}
where $\phi = N(\sigma/(2R))^3$ is the monomer volume fraction.  This
result can be understood within blob scaling theory by realizing that,
unlike the planar or cylindrical case, the monomer concentration within
a spherical cavity increases with increasing polymer size~$N$.  The
extensivity of the free energy of confinement is then recovered only
when a change in polymer size is accompanied by a corresponding change
in volume of the cavity such that the monomer concentration remains
invariant~\cite{grosberg94,sakaue06}.

\begin{figure}[ht]
  \begin{center}
  \includegraphics*[width=0.20\textwidth]{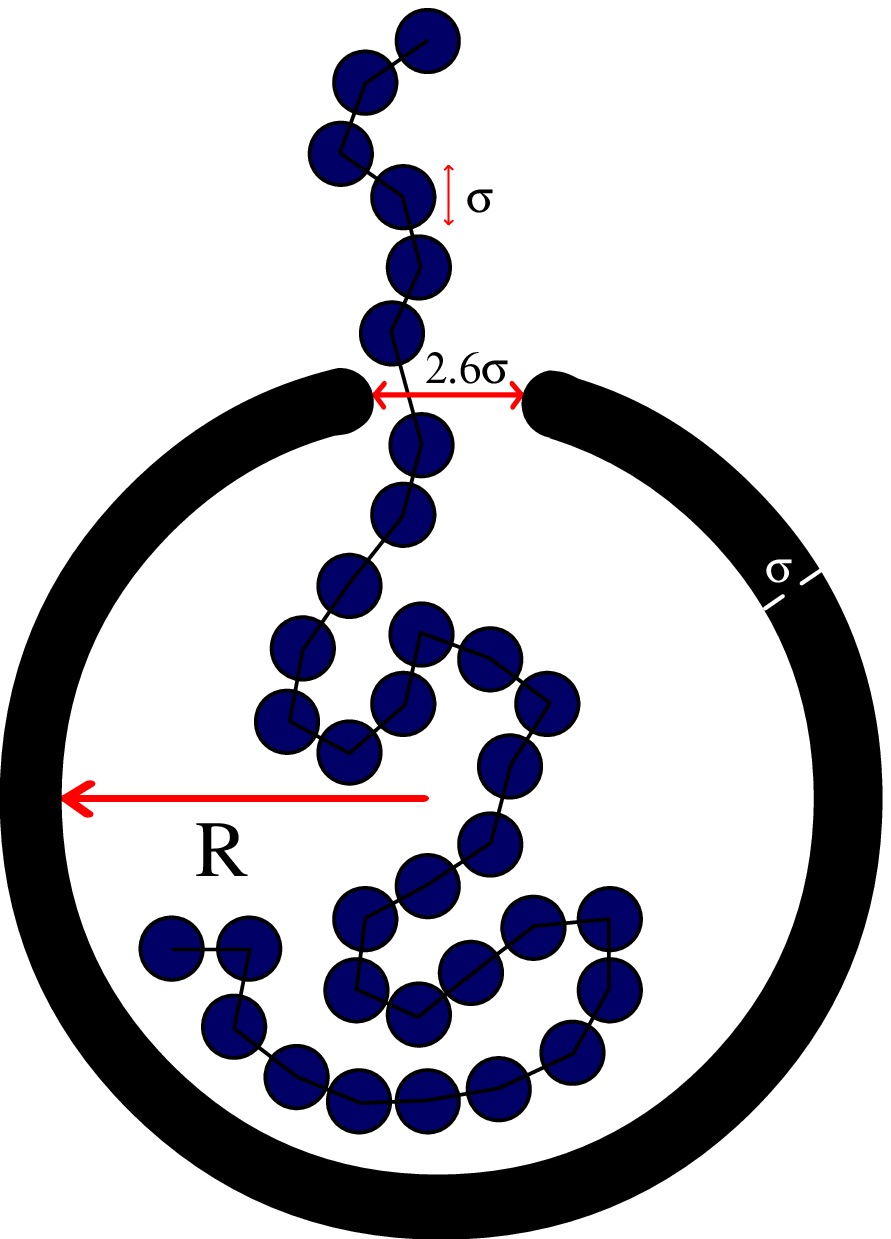}
  \hspace*{2mm}
  \includegraphics*[width=0.22\textwidth]{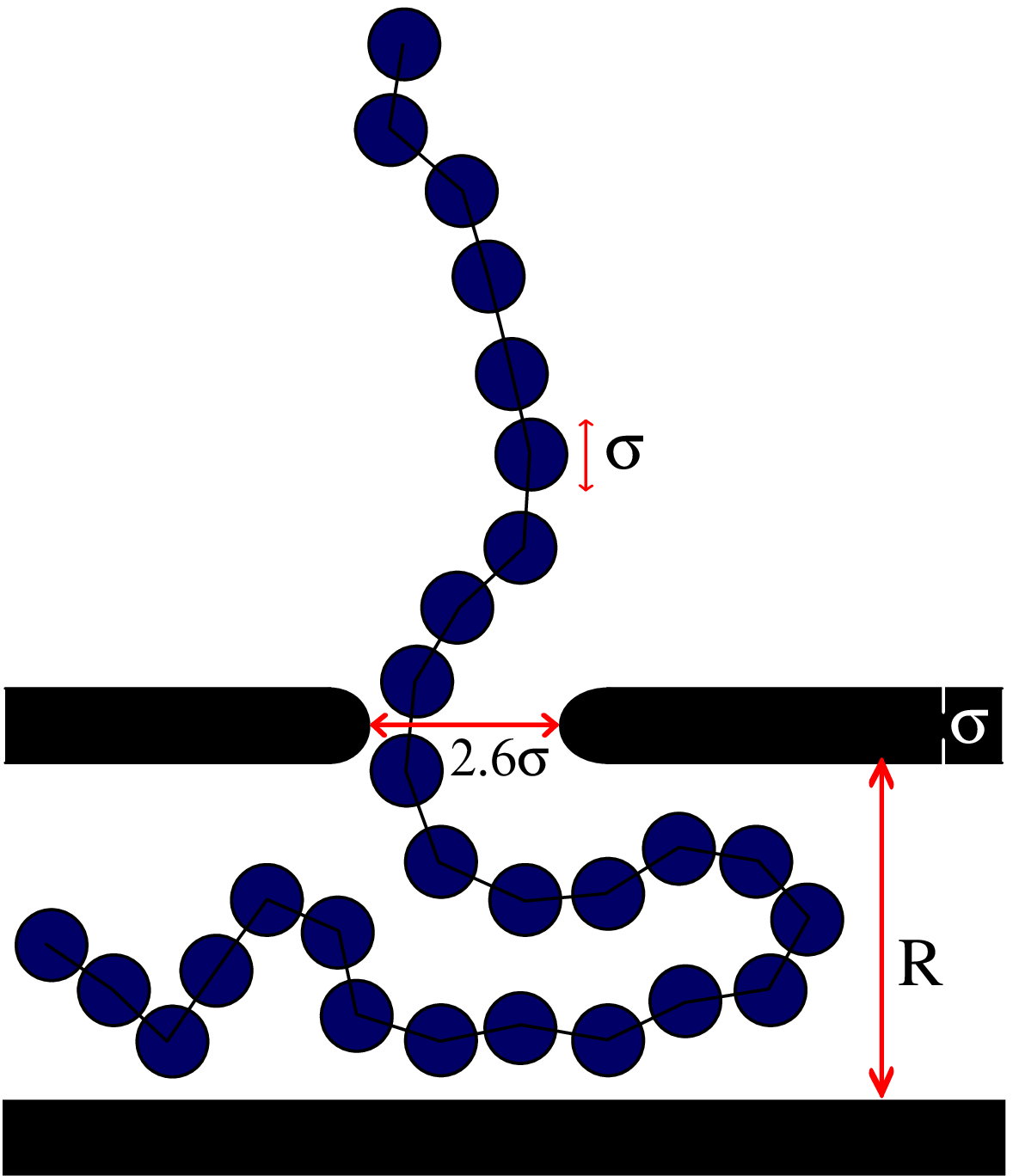}
  \end{center}
  \caption{Schematic setup of the Monte Carlo simulations. In the
  left-hand panel, a polymer chain is released from a spherical geometry
  of radius~$R$. In the right-hand panel, the chain is confined between
  two parallel plates at separation~$R$ and escapes via a circular
  opening in one of the plates. The actual simulations are performed in
  three dimensions.}
  \label{fig:image}
\end{figure}

Since the free energy of confinement per monomer, $\Delta F/N$,
represents the chemical potential gradient that drives the
translocation, combination of Eqs.\ (\ref{eq:plates})
and~(\ref{eq:sphere}) with Eq.~(\ref{eq:escapetime}) leads to
predictions for the translocation out of a planar or a spherical
geometry, respectively,
\begin{subnumcases}{\tau \sim \label{eq:dynamics}}
  N^{1+\nu} \left( \frac{R}{\sigma} \right)^{1/\nu}
  & planar confinement     \label{eq:dynamics-planes} \\
  N^{1+\nu} \phi^{1/(1-3\nu)}
  & spherical confinement  \label{eq:dynamics-sphere}
\end{subnumcases}
where the exponent $1+\nu$ in the prefactor represents the lower bound
proposed in Ref.~\cite{kantor04}.
To validate this prediction, we study the translocation of a flexible
polymer chain which is modeled as a linear series of $N$ spherical beads
of diameter $\sigma$, connected by bonds that are freely extensible up
to a fixed value $\ell_{\rm M}$. All monomers interact via a hard-core
repulsion,
\begin{equation}
u_m(r_{ij}) = \left\{
\begin{array}{ll}
  0      & \quad  r_{ij}    > \sigma \\
  \infty & \quad  r_{ij}    \leq \sigma
\end{array}
\right. \;,
\end{equation}
where $r_{ij}$ is the center-to-center distance between beads $i$
and~$j$. The nearest-neighbor bonds are represented by
\begin{equation}
u_b(r_{i,i-1}) = \left\{
\begin{array}{l l}
      0      & \quad  r_{i,i-1} \leq \ell_{\rm M} \\
      \infty & \quad  r_{i,i-1}    > \ell_{\rm M}
\end{array}
\right. \;.
\end{equation}
We mimic the dynamical properties of this model by means of Monte Carlo
simulations in which only local, short-ranged displacements are
employed. The monomer displacement per Monte Carlo step equals $(\Delta
x, \Delta y, \Delta z)$, in which each cartesian component is chosen
uniformly in the range $[-0.15\sigma, 0.15\sigma]$~\cite{note-isotropy}.
To avoid dynamical inconsistencies that could result from crossing
polymer bonds, we choose $\ell_{\rm M} = \smash{\sqrt2} \sigma$.
Confinement is imposed by means of a spherical or planar boundary of
thickness~$\sigma$, which exerts a hard-core repulsion on the monomers.
For the planar case, the system is periodically replicated in the
directions parallel to the plates, with a period $2N\sigma$. The setup
is depicted schematically in Fig.~\ref{fig:image}.

The polymer chain is first equilibrated within the confining geometry.
Subsequently, we create a smooth pore (shaped as the hole in a torus) of
radius $1.3\sigma$ to allow ejection of the polymer (see
Fig.~\ref{fig:image}).  This radius is sufficiently small to practically
exclude the translocation of folded chains.  A translocation event is
considered successful when the entire polymer escapes from the confining
geometry.  We define the translocation time~$\tau$ as the difference
between the time~$t_N$ when the last monomer has left the cavity and the
time~$t_1$ when, within a successful event, the first monomer has
entered the pore.  We systematically vary the degree of confinement~$R$
and determine the average escape time for chains with lengths ranging
from $N=40$ to $N=512$ monomers. For each choice of $R$ and $N$, we
generate between 400 and 1400 independent translocation events.  All
simulations are performed in the drift-dominated regime, i.e.,
$N\Delta\mu > 1$, which requires $R < R_G$.

\begin{figure}[bt]
  \centerline{\includegraphics*[width=\figurewidth]{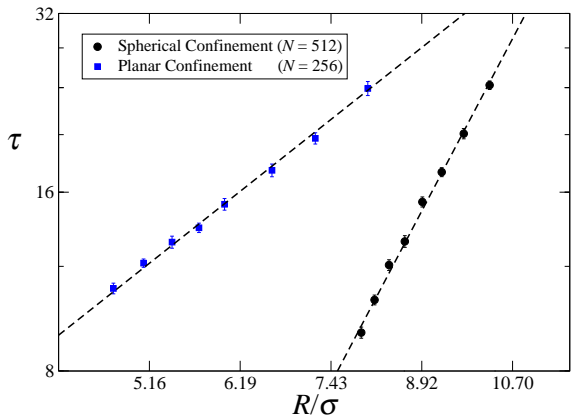}}
  \caption{Double logarithmic plot of the average translocation time
  $\tau$ for a linear, flexible polymer escaping from a spherical and a
  planar geometry, as a function of the degree of confinement~$R$.  The
  translocation times are expressed in units of $10^6$ Monte Carlo
  sweeps. These data confirm the striking dependence on confinement
  geometry, Eqs.\ (\ref{eq:dynamics-planes})
  and~(\ref{eq:dynamics-sphere}).}
  % Least-squares fits (dashed line) yield a power-law dependence $\tau
  % = \tau_0 + aR^\gamma$ in good agreement with the predictions of
  % Eqs.\ (\ref{eq:dynamics-planes}) and~(\ref{eq:dynamics-sphere}).}
  \label{fig:tau_vs_R}
\end{figure}

To focus on the role of the degree of confinement, we calculate $\tau$
for a fixed, long chain length ($N=256$ for planar and $N=512$ for
spherical confinement). Figure~\ref{fig:tau_vs_R} displays $\tau$ as a
function of~$R$. The escape times from both geometries are accurately
described by a power-law dependence, but with strikingly different
exponents.  For confinement within a planar geometry, the driving force
is relatively weak and we use smaller separations than for the spherical
geometry (as small as $R=4.8\sigma$). A least-squares fit of both data
sets to the expression $\tau(R) = \tau_0 + a R^{\gamma}$ yields
$\gamma=1.54 \pm 0.10$ for planar confinement and $\gamma=3.65 \pm 0.08$
for spherical confinement, with chi-square per degree of freedom
($\bar{\chi}^2$) equal to $1.06$ and $0.86$, respectively.  These
results are in good agreement with the exponents in $1/\nu$
[Eq.~(\ref{eq:plates})] and $3/(3\nu-1)$ [Eq.~(\ref{eq:sphere})] and
thus confirm the linear dependence of $\tau$ on $1/\Delta\mu$ predicted
by Eq.~(\ref{eq:dynamics}). The agreement is even closer if one notes
that for the chain lengths employed here the effective Flory exponent is
slightly larger than $\nu =0.588$.

\begin{figure}[ht]
  \centerline{\includegraphics*[width=\figurewidth]{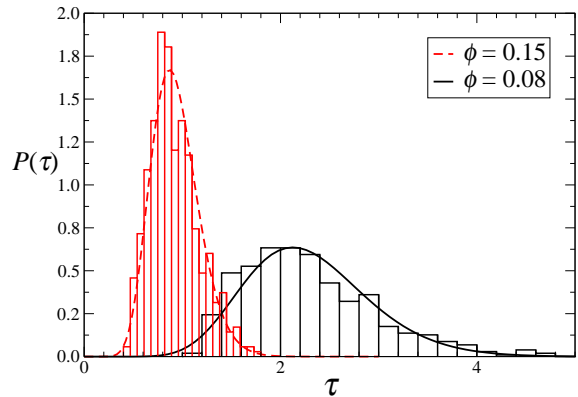}}
  \caption{Probability distribution~$P(\tau)$ of translocation
  times~$\tau$, for a polymer of $N=128$ monomers confined in a
  spherical cavity. The translocation times are expressed in units of
  $10^6$ Monte Carlo sweeps.  The narrow distribution corresponds to a
  strongly confined chain (monomer volume fraction $\phi=0.15$) and the
  wider distribution is obtained for a smaller driving force
  ($\phi=0.08$).  As discussed in the text, the curves represent
  empirical expressions of the form $\tau^{a_1} \exp(-a_2 \tau)$, with a
  characteristic ratio $\Delta$ of the width and the peak position that
  is close to typical experimental results.}
  \label{fig:Pdistrib}
\end{figure}

Additional information about the translocation process can be obtained
from the probability distribution function of translocation times
$P(\tau)$. % (i.e., the distribution of passage times).
Theoretical~\cite{lubensky99} and experimental
results~\cite{kasianowicz96,meller01,chen04} indicate that this
distribution deviates from a Gaussian distribution and may be
considerably skewed. Consequently, the \emph{average} translocation time
is not fully representative of the experimental data.  We sample
$P(\tau)$ for a chain of $N=128$ monomers, comparable to the chain
length employed in the experimental determination of $P(\tau)$ for
single-stranded DNA~\cite{kasianowicz96}.  For escape from a spherical
cavity, the driving force depends on the monomer volume fraction.  As
shown in Fig.~\ref{fig:Pdistrib}, for strong confinement ($\phi=0.15$),
the distribution is narrow and nearly Gaussian. However, at weaker
confinement ($\phi=0.08)$, $P(\tau)$ broadens and the skewness becomes
clearly visible. Lubensky and Nelson~\cite{lubensky99} derived an
expression that provides a reasonable description of the data in
Fig.~\ref{fig:Pdistrib}; however, this expression is not valid for
large~$\tau$.  Following the experimental analysis~\cite{meller01} we
therefore fit the data to an empirical expression of the form
$\tau^{a_1} \exp(-a_2 \tau)$.  Just as in the experiments, the
exponential term provides a good description of the long-time tail. It
is argued in Ref.~\cite{lubensky99} that the distribution of passage
times can be characterized in a useful way via the ratio $\Delta \equiv
\delta\tau/\tau_{\rm max}$ between the width $\delta\tau$ of the
distribution (as defined in Ref.~\cite{lubensky99}) and its peak
position~$\tau_{\rm max}$. For the distributions shown in
Fig.~\ref{fig:Pdistrib} we find values in the range 0.53--0.56, indeed
in agreement with experimental values $\Delta \approx
0.5$~\cite{kasianowicz96,lubensky99} and $\Delta \approx
0.55$~\cite{meller01}.

\begin{figure}
  \centerline{\includegraphics*[width=\maximumfigurewidth]{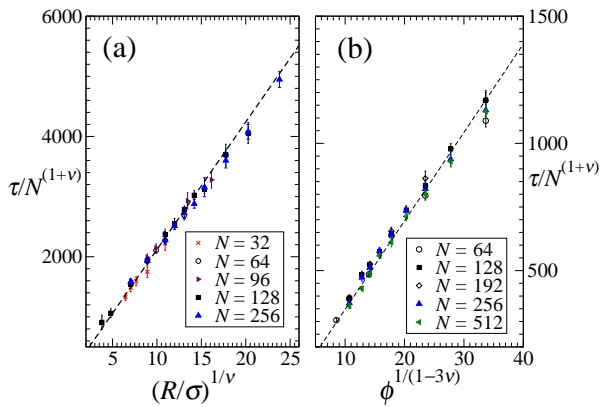}}
  \caption{Average translocation time $\tau$ (in units of Monte Carlo
  sweeps) normalized by $N^{1+\nu}$ for chains escaping from (a) planar
  confinement, as a function of the inverse driving force per monomer
  $(\Delta\mu)^{-1} \sim (R/\sigma)^{1/\nu}$, and (b) spherical
  confinement, as a function of $(\Delta\mu)^{-1} \sim
  \phi^{1/(1-3\nu)}$. The collapse of the data points for different
  chain lengths confirms the validity of Eqs.\ 
  (\ref{eq:dynamics-planes}) and~(\ref{eq:dynamics-sphere}),
  respectively.}
  \label{fig:collapseALL}
\end{figure}

We now proceed to determine the dependence of translocation time on the
degree of polymerization~$N$. For the spherical case, we perform a
series of simulations at constant initial volume fraction $\phi=0.1$,
and for the planar geometry we perform a series of simulations at fixed
separation $R=4.8\sigma$. For both geometries $\tau$ is accurately
described by a power-law dependence $\tau \sim N^\delta$ that is
independent of geometry, in accordance with the
observation~\cite{grosberg94,cacciuto06} that the free energy of
confinement is extensive [cf.\ Eqs.\ (\ref{eq:plates})
and~(\ref{eq:sphere})].  For planar confinement, a least-squares fit
yields $\delta=1.55 \pm 0.04$ ($\bar{\chi}^2=1.16$) and for spherical
confinement we find $\delta=1.59 \pm 0.03$ ($\bar{\chi}^2=0.87$). Both
results agree with $1+\nu\approx 1.59$, confirming the lower bound
established by Eq.~(\ref{eq:escapetime}).  Accordingly, all results for
a given geometry can be combined in a single data collapse.
Figure~\ref{fig:collapseALL}(a) (plates) shows the translocation time
$\tau$ normalized by $N^{1+\nu}$ as a function of the inverse driving
force $(\Delta\mu)^{-1} \sim (R/\sigma)^{1/\nu}$, for five different
chain lengths ($N=32$, $64$, $96$, $128$, and $256$). Likewise,
Fig.~\ref{fig:collapseALL}(b) (sphere) displays $\tau/N^{1+\nu}$ as a
function of $\Delta\mu^{-1}\sim \phi^{1/(1-3\nu)}$ for $N=64$, $128$,
$192$, $256$, and $N=512$.  In both cases, all data are described by a
single master curve with $\bar{\chi}^2=1.40$ and $1.45$, respectively.

In view of the striking difference between planar and spherical geometry
that we observe for the $R$-dependence of the translocation time
(Fig.~\ref{fig:tau_vs_R}), it is remarkable that earlier
work~\cite{muthukumar01} found good agreement between numerical results
for \emph{spherical} confinement and a theoretical prediction based upon
Eq.~(\ref{eq:escapetimeLinear}) and Eq.~(\ref{eq:plates}), i.e., a
driving force obtained from the free energy of \emph{uniaxial} (i.e.,
planar) confinement. We ascribe this finding to the fortuitous
cancellation of two errors. Indeed, in Ref.~\cite{muthukumar01} the
translocation time is predicted to scale as
\begin{equation}
  \tau \sim N \left( \frac{\sigma}{R} \right)^{1/\nu} 
       \sim N \left( \frac{N}{\phi} \right)^{1/(3\nu)} \;.
\label{eq:tau-muthukumar}
\end{equation}
Combination of the linear $N$ dependence of
Eq.~(\ref{eq:escapetimeLinear}) and the inappropriate expression for the
free energy of confinement yields an overall chain-length dependence
$N^{1+1/(3\nu)}$, so that, at fixed concentration, $\tau$ is predicted
to scale as $N^{1.567}$, which coincidentally is in approximate
numerical agreement with the lower bound $\tau \sim N^{1+\nu}$
[Eq.~(\ref{eq:escapetime})]. In fact, since the data in
Ref.~\cite{muthukumar01} exhibit a collapse when scaled by the $N$
dependence of Eq.~(\ref{eq:tau-muthukumar}) we conclude that those data
corroborate our findings.  It is more difficult to reconcile our
findings in Fig.~\ref{fig:tau_vs_R} with the apparent confirmation of
the concentration dependence of Eq.~(\ref{eq:tau-muthukumar}).  However,
we note that the evidence in Ref.~\cite{muthukumar01} consists of a
scaling collapse rather than an explicit numerical analysis of the power
law. When performed over a narrow range of densities, such a collapse
can be relatively insensitive to the precise power of~$\phi$.

In summary, we have investigated the translocation of a flexible polymer
chain through a hole, when the driving force is generated by confinement
of the polymer. To clarify the role of the confinement geometry, we have
considered the escape of a polymer from a planar as well as a spherical
geometry. For both cases, we demonstrate that the translocation time has
a chain-length dependence that follows the nonlinear lower bound
established by Kantor and Kardar~\cite{kantor04}. The driving force
affects the translocation time via a \emph{linear} dependence on the
inverse chemical potential gradient, as predicted on analytical
grounds~\cite{sung96,muthukumar99}. Thus, the average translocation time
can be described by a simple scaling relation, $\tau \sim
N^{1+\nu}/\Delta\mu$, which results in a \emph{geometry-dependent} power
law when expressed in terms of the length scale of
confinement---distance between parallel plates for a planar geometry and
cavity radius for a spherical geometry.

\begin{acknowledgments}
  We acknowledge helpful discussions with Lei Guo.  This material is
  based upon work supported by the U.S. Department of Energy, Division
  of Materials Sciences under Award No.\ DEFG02-91ER45439, through the
  Frederick Seitz Materials Research Laboratory at the University of
  Illinois at Urbana-Champaign.  We also acknowledge computing time on
  the Turing Xserve Cluster at the University of Illinois.
\end{acknowledgments}

%\bibliographystyle{apsrev}
%\bibliographystyle{prsty}
%\bibliography{journals,translocation,polymer,conf-trans-notes}

\begin{thebibliography}{10}

\bibitem{alberts02}
B. Alberts {\it et~al.}, {\em Molecular Biology of the Cell}, 4th ed. (Garland
  Science, New York, N.Y., 2002).

\bibitem{kasianowicz96}
J.~J. Kasianowicz, E. Brandin, D. Branton, and D.~W. Deamer, Proc. Natl. Acad.
  Sci. U.S.A. {\bf 93},  13770  (1996).

\bibitem{akeson99}
M. Akeson {\it et~al.}, Biophys. J. {\bf 77},  3227  (1999).

\bibitem{meller00}
A. Meller {\it et~al.}, Proc. Natl. Acad. Sci. U.S.A. {\bf 97},  1079  (2000).

\bibitem{sung96}
W. Sung and P.~J. Park, Phys. Rev. Lett. {\bf 77},  783  (1996).

\bibitem{park98b}
P.~J. Park and W. Sung, Phys. Rev. E {\bf 57},  730  (1998).

\bibitem{lubensky99}
D.~K. Lubensky and D.~R. Nelson, Biophys. J. {\bf 77},  1824  (1999).

\bibitem{muthukumar99}
M. Muthukumar, J. Chem. Phys. {\bf 111},  10371  (1999).

\bibitem{muthukumar01}
M. Muthukumar, Phys. Rev. Lett. {\bf 86},  3188  (2001).

\bibitem{chuang01}
J. Chuang, Y. Kantor, and M. Kardar, Phys. Rev. E {\bf 65},  011802  (2001).

\bibitem{tian03}
P. Tian and G.~D. Smith, J. Chem. Phys. {\bf 119},  11475  (2003).

\bibitem{lansac04}
Y. Lansac, P.~K. Maiti, and M.~A. Glaser, Polymer {\bf 45},  3099  (2004).

\bibitem{kantor04}
Y. Kantor and M. Kardar, Phys. Rev. E {\bf 69},  021806  (2004).

\bibitem{kong04}
C.~Y. Kong and M. Muthukumar, J. Chem. Phys. {\bf 120},  3460  (2004).

\bibitem{storm05}
A.~J. Storm {\it et~al.}, Nano Letters {\bf 5},  1193  (2005).

\bibitem{meller01}
A. Meller, L. Nivon, and D. Branton, Phys. Rev. Lett. {\bf 86},  3435  (2001).

\bibitem{chen04}
P. Chen {\it et~al.}, Nano Letters {\bf 4},  2293  (2004).

\bibitem{grosberg94}
A.~Y. Grosberg and A.~R. Khokhlov, {\em Statistical Physics of Macromolecules}
  (American Institute of Physics, New York, N.Y., 1994).

\bibitem{cacciuto06}
A. Cacciuto and E. Luijten, Nano Letters {\bf 6},  901  (2006).

\bibitem{degennes79}
P.-G. de~Gennes, {\em Scaling Concepts in Polymer Physics} (Cornell University
  Press, Ithaca, N.Y., 1979).

\bibitem{sakaue06}
T. Sakaue and E. Rapha{\"e}l, Macromolecules {\bf 39},  2621  (2006).

\bibitem{note-isotropy}
The anisotropy in this choice has, we have verified, a negligible effect on the
  translocation times.

\end{thebibliography}

\end{document}